\documentclass[notitlepage,12pt,onecolumn,showpacs,amssymb,aps,nofootinbib,floatfix]{revtex4-1}
\usepackage{epsfig,color,bigints}

\renewcommand{\eqref}[1]{Eq. \ref{#1} }

\newcommand{\eqcomma}{\phantom{AA},\phantom{AA}}

\newcommand{\lnz}{\ln \mathcal{Z}}

\newcommand{\order}[1]{ \mathcal{O} \left( #1 \right) }

\begin{document}
\title{Emergent symmetries of relativistic fluid dynamics from local ergodicity}
\begin{abstract}
  We show that volume-preserving diffeomorphisms and the chemical shift symmetry defining relativistic lagrangian ideal fluid dynamics can be derived as an emerging symmetry when ergodicity is assumed to apply locally in a way that is invariant under smooth spacetime foliations.   This can be used as a way to derive the ideal hydrodynamic limit in a strongly coupled but strongly fluctuating medium.
  We comment on the connection with thermalization in small systems,
  the Eigenstate thermalization hypothesis and deviations from the ideal limit.
\end{abstract}
\author{Giorgio Torrieri$^1$}
\affiliation{$\phantom{A}^1$Universidade Estadual de Campinas - Instituto de Fisica "Gleb Wataghin"\\
Rua Sérgio Buarque de Holanda, 777\\
 CEP 13083-859 - Campinas SP\\
torrieri@ifi.unicamp.br\\
}
\maketitle
%\section{Classical ergodicity}
Relativistic hydrodynamics has proven to be an extremely useful phenomenological tool \cite{kodama,rom,heinz} in the study of heavy ion collisions.
This has also reignited the more formal theoretical study of hydrodynamics \cite{denicol,hydeft,zubecc}, a subject much more complex and subtle on a conceptual level than one initially thinks.  The observation of hydrodynamic behavior in ``small'' systems, both in size and number of degrees of freedom \cite{zajc} (but also analogous non-relativistic examples such as ultracold atoms \cite{ultracold} and the ``Brazil-nut effect'' \cite{braznut}) demonstrates that conceptual ambiguities can still be relevant for resolving mysteries given to us by experimental data.

One such conceptual issue is that there are several definitions to hydrodynamics:  One usually thinks of it as a coarse-grained, long range effective theory written in terms of conserved charges and their gradients \cite{kodama,rom,denicol}.   In this approach, hydrodynamic evolution is explicitly deterministic, and the connection to local thermal equilibrium is not ``built in'', leading to concepts such as hydrodynamic attractors far from thermal equilibrium \cite{attractor,mrow}.   The potential ambiguity comes from the fact  that methods from statistical mechanics are necessary to calculate the equation of state and transport coefficients needed to close the hydrodynamic system of equations.

Statistical mechanics however is not a deterministic theory, at best fluctuations vanish in the thermodynamic limit, but in principle it's only assumption is the equal probability of microstates\cite{khinchin,jaynes}.    One therefore needs to make sure fluctuations stay ``small'', something usually implicitly assumed in microscopic theories rather than demonstrated \cite{ll,gale,csernai,kovfluct,kapusta,teaney,jorgefluct,sticky,sk1,sk2,steph} (Kubo formulae, for instance, are based on taking limits much larger than the fluctuation timescale \cite{heinz}), though fluctuation-dissipation relations apply also away from the thermodynamic limit \cite{zubarev,bonanca}.

As a perhaps related issue, away from perfect equilibrium the definition of ``equilibrium'' becomes itself ambiguous \cite{kovtun,spalinski,prx,gavassino,gava,bdnk}.  A full definition of hydrodynamics in this sense, as well as a connection to non-equilibrium quantum field theory that encodes microscopic fluctuations and their hydrodynamic response \cite{arnold,gelis,calzetta},  is still missing.

A still different definition of hydrodynamics is based on a lagrangian picture, where instead of defining currents in the lab frame we follow every cell of the fluid as it propagates and deforms.   This definition is particularly elegant in that ideal hydrodynamics is based on local diffeomorphisms \cite{jackiw,nicolis,son}, making it into a  ``poor people's general relativity'' \cite{hydeft,analogue} (to the extent that perhaps the problem of quantizing gravity can be related, via the enthropic hypothesis \cite{jacobson}, to the problem, described in the previous paragraph, of defining a fluctuating hydrodynamics where local symmetries still hold \cite{zubarev,deco}).

This work makes a link between this last Lagrangian definition of hydrodynamics and the previous two, by showing it's elegant local symmetries are in fact emerging if one assumes the ergodic hypothesis (Defined via \eqref{ergodef} rather than the usual $\Delta \rightarrow \infty$ limit \cite{khinchin}) applies in every cell of the fluid.
This link can then be used to define hydrodynamics in a way that might be more appropriate to small systems \cite{zajc}, and as a starting point to examine the fluctuations and redundancies of \cite{bdnk}.

Statistical mechanics is based on the emergence of an approximately ergodic system\cite{ll} (or rather \cite{khinchin} of the approximation that some observables are described by such a system), whose evolution is governed by an equation of the type
\begin{equation}
\label{ergodef}
  \left. \frac{\Delta x}{\Delta t}\right|_{t-t'\simeq \Delta t } \simeq \int P(q')dq'
\end{equation}
where $x$ is a microscopic phase space degree of freedom, $t$ is a time sampled across some average interval $\Delta t$,and $dq'$ is a time-invariant measure over an ensemble of states.

In other words when sampled under a certain coarse-graining timescale, the system's ``evolution'' can be approximated via a 
statistical distribution of microstates.  As shown in \cite{jaynes,khinchin}, while this is an extremely fragile assumption, dependent 
on the "indecomposability" of the phase space hypersurface constrained by conservation laws, it can be thought of as an approximate 
assumption good enough to calculate particular observables (in particular conserved currents) and their ''typical'' state in strongly 
coupled systems whose initial conditions are sampled across a broad ensemble.   We also note that typically the ergoic hypothesis is
formulate for $t-t' \rightarrow \infty$, while here this interval defines the size of some cell.  This is in line with thermalization going
from an asymptotic quantity to a local one in the definition of fluid mechanics.

Thus, for a physical system moving in phase space sufficiently chaotically, with  conservation laws of energy and momentum ($P^\mu$ and their invariant $P$) as well as charge $Q$ but not necessarily particle number
\begin{equation}
  \label{ergdef}
  \left. \frac{\Delta \phi }{\Delta t}\right|_{t-t'=\Delta t} \simeq \frac{1}{d\Omega(Q,P^\mu)} \sum \delta^4_{P^\mu,P_{macro}^\mu(t)}\delta_{Q,Q_{macro}(t)}  \delta\left( \sum_j^{\infty} p_j^\mu- P^\mu    \right) \delta\left( \sum_j^{\infty} Q_j- Q    \right)
\end{equation}
where $\phi=\{ x_{1,2,....N\rightarrow \infty}  \}$ is some degree of freedom, $p_i$ is the conjugate momentum and $Q_i$ some conserved charge and $d \Omega(P^\mu,Q)$ is the normalization for that point in spacetime, ( $\sim e^{dS}$ where $dS$ is the entropy density of the cell).
Note that for varying particle numbers the Gibbs factor $N!$ can be taken to be a correction of $d\Omega$, so we do not consider it further.
From this the equilibrium state in the micro-canonical ensemble. Other ensembles  follow when the number of degrees of freedom and volume are taken to infinity.

It is immediately clear that the ergodic hypothesis is not relativistically covariant, since there is a ``t'' coordinate on the left hand side (the right hand side can be made generally covariant by an on-shell condition).  For ``global'' equilibrium, this is of course not a problem: The system as a whole carries momentum, so there is a preferred frame where it's momentum is zero and hence only the energy constraint is needed to maximize entropy.

But this seems in contradiction \cite{gava} with the concept of ``local'' equilibrium, where every cell of a fluid contains ``many'' degrees of freedom in ergodic equilibrium.     To an extent, this is not a deep issue since local equilibrium is not rigorously defined beyond the ideal limit.   However, it would be interesting to try to define it from the ergodic hypothesis and approximate deviations from it.
To do so, we would like to investigate what kind of symmetries are needed to be imposed on Eq. \ref{ergdef} for a relativistic local ergodic evolution to emerge.

To try to add an element of covariance, we define ``t'' via a foliation, $\Sigma_\mu$\footnote{Here we use the notation first introduced by Cooper and Frye in their treatment of Freezeout \cite{cf} and further used in Zubarev hydrodynamics \cite{zubecc}, where a "space-like" volume element can be parametrized by a time-like 4-vector denoted by $\Sigma_mu$ whose exterior derivative points to the 4-volume's proper time.    This notation is a bit different than the one usually used in general relativity \cite{grfol}, more appropriate for curved spacetimes }.   
Different $\Sigma_\mu$'s are related via generally non-inertial transformations.   We would like to concentrate on transformations that respect the causality of the foliation
\begin{equation}
  \label{smooth}
d\Sigma_\mu = \frac{\partial \Sigma_\mu}{\partial \Sigma'_\nu}d\Sigma'_\mu \eqcomma \underbrace{(d \Sigma)^2}_{d\tau^2},\underbrace{( d\Sigma')^2}_{d\tau'^2} >0 \eqcomma  \left. \frac{d}{dt} \right|_{dt\sim \Delta t} \rightarrow \left. \frac{d}{d\Sigma_0} \right|_{d \tau \sim \Delta t} \eqcomma d\Sigma_\mu \det \left(\partial_\mu \Sigma_\nu\right)^{-1}\leq \Delta
\end{equation}
A definition of ``ideal'' local equilibrium is to ensure that Eq. \ref{ergdef} stays true independently of $\Sigma_\mu$ as long as $\Sigma$ is causal.  In this picture, the averaging is done not on $dt \sim \Delta t$ but on $d\tau \sim \Delta t$, for each cell.  The change in foliation must be smooth enough as to not disturb the average $\Delta t$ (the ``microscopic scale'', which in this case is the maximum allowed radius of curvature of the foliations).

To see if this is possible,
let us remember that $p_\mu$ and $Q$ are Noether currents, corresponding to symmetries of the Lagrangian
\begin{equation}
  \label{pdef}
  p_\mu=\int d^3 \Sigma^\nu T_{\mu \nu} \eqcomma T_{\mu \nu}=  \frac{\partial L}{\partial {\partial^\mu \phi}} \Delta_\nu \phi - g_{\mu \nu}L  \eqcomma \Delta_\nu \phi(x_\mu)=\phi(x_\mu+dx_\nu)
\end{equation}
\begin{equation}
  \label{qdef}
  Q=\int d^3 \Sigma^\nu j_{ \nu} \eqcomma j_{ \nu}=  \frac{\partial L}{\partial {\partial^\mu \phi}} \Delta_\psi \phi \eqcomma \Delta_\psi\phi= \phi(x_\mu)+\delta \phi^* (x_\mu) = |\phi(x)|e^{ i \left( \psi(x)+\delta \psi(x) \right) }
\end{equation}
and $\psi=\{ \psi_{1,2,...N\rightarrow \infty}  \}$ is a complex phase.

These textbook formulae provide us with an intuition of how to model local equilibrium.   All one has to do is to impose ergodicity on every space-time point and then make this hypothesis foliation-independent.
Combining Eq. \ref{ergdef} with \ref{pdef} and \ref{qdef} we get
\begin{equation}
  \label{ergdelta}
  \left. \frac{\Delta \phi }{\Delta \Sigma_0}\right|_{t-t'=\Delta t} = \frac{1}{d\Omega(dP^\mu(\Sigma_0),dQ(\Sigma_0))} \sum \delta^4 \left(  d\Sigma^\nu \underbrace{ \frac{\partial L}{\partial {\partial^\mu \phi}} \partial_\nu \phi- L d\Sigma^\mu}_{\delta L (a_\mu)  \sim a_\alpha \partial^\alpha \left( \delta_\nu^\mu L\right)}-  dP^\mu(\Sigma_0)    \right) \times
\end{equation}
\[\ 
\times  \delta\left(  d\Sigma^\mu \underbrace{ \frac{\partial L}{\partial {\partial^\mu \phi}} \frac{\partial \psi}{\partial \Sigma_0}}_{\delta L(\psi)\sim \psi \partial_\mu L  } - dQ(\Sigma_0)   \right)
\]
and enforce the covariance of these equations with any change in foliation that respects causality ($d\Omega,dP,dQ$ refer to the phase space volume, momentum and charge within the cell).
Let us therefore use $\delta a_\mu$ and $\delta \psi$ as coarse-graining degrees of freedom for a generic equation of motion. 
\begin{equation}
  \label{ergdelta2}
  \left. \frac{\Delta \phi }{\Delta \Sigma_0}\right|_{t-t'=\Delta t} = \frac{1}{d\Omega(P^\mu,Q,\Sigma_0)} \int da_\mu d\psi \delta^4 \left(   d\Sigma^\nu a_\alpha \partial^\alpha \left( \delta_\nu^\mu L\right)-  dP^\mu(\Sigma_0)    \right) \delta\left( d\Sigma^\mu  \psi \partial_\mu L   - dQ(\Sigma_0)    \right)
\end{equation}
Let us now re-write Eq. \ref{ergdelta2} under a different foliation, and see what kind of constraints will foliation invariance of the ergodic hypothesis bring.
Provided both foliations are smooth enough (According to Eq. \ref{smooth}) the LHS simply becomes
\begin{equation}
 \left. \frac{\Delta \phi }{\Delta t}\right|_{t-t'=\Delta t} \rightarrow \frac{d \Sigma'_\nu}{d \Sigma_0} \left. \frac{\Delta \phi }{\Delta \Sigma_0} \right|_{\Sigma_0-\Sigma_0'=\Delta t} = \frac{\partial \Sigma'_\nu}{\partial \Sigma_0} \left. \frac{\Delta \phi }{\Delta \Sigma_\nu} \right|_{x_\mu -x_\mu'<\Delta t} 
\label{ergfol}
\end{equation}
now we eliminate the LHS from the system of equations of \eqref{ergdelta2} in $\Sigma_\mu$ and $\Sigma'_\mu$. The resulting equation, a ratio of the RHSs, will be
\begin{equation}
  \label{ergdeltashift}
\frac{d\Omega (dP_\mu',dQ',\Sigma_0')}{d\Omega (dP_\mu,dQ,\Sigma_0)}= \frac{d \Sigma'_0}{d \Sigma_0}
 \frac{ \int da_\mu d\psi \delta^4 \left(   d\Sigma^\nu a_\alpha \partial^\alpha \left( \delta_\nu^\mu L\right)-  dP^\mu(\Sigma_0)    \right) \delta\left( d\Sigma^\mu  \psi \partial_\mu L   - dQ(\Sigma_0)    \right)}{\int da'_\mu d\psi' \delta^4 \left(   d\Sigma'_\nu a_\alpha' \partial^\alpha \left( \delta_\nu^\mu L\right)-  dP_\mu'(\Sigma_0')    \right) \delta\left( d\Sigma_\mu'  \psi' \partial^\mu L   - dQ'(\Sigma_0')    \right)}
\end{equation}
 %of Eq. \ref{ergdelta} into space and time componets, the RHS becomes\[\   \int  \delta^4 \left( d\Sigma^0 \frac{\partial L}{\partial {\partial^\mu \phi}} \dot{\phi}  - d\Sigma^i \frac{\partial L}{\partial {\partial^\mu \phi}} \partial_i \phi- L d\Sigma^\mu-  P^\mu     \right)  \delta\left( \int d\Sigma^\mu  \frac{\partial L}{\partial {\partial^\mu \phi}} \frac{\partial \psi}{\partial \Sigma_0} - Q    \right)   \]
We also take advantage of the fact that $\Sigma_\mu$ are causal, an parametrize $d \Sigma_\mu$ by three Cartesian coordinates $\Phi_{I=1,2,3}$ such that the tangent vector
\begin{equation}
  \label{localfoliation}
d\Sigma_\mu =  \epsilon_{\mu \nu \alpha \beta} \frac{\partial \Sigma^\nu}{\partial \Phi_1} \frac{\partial \Sigma^\alpha}{\partial \Phi_2} \frac{\partial \Sigma^\beta}{\partial \Phi_3} d\Phi_1 d\Phi_2 d\Phi_3  
\end{equation}
where the latin letters are 1,2,3 and 0 would be the time coordinate is a time-like 3-volume element for a certain normalization.   Arbitrary changes in foliation can also be decomposed in
\begin{equation}
\frac{\partial \Sigma'_\mu}{\partial \Sigma_\nu}=\Lambda_\mu^\nu \det \frac{d\Phi'_I}{d\Phi_J} \eqcomma \det \Lambda_{\mu}^\nu=1
\end{equation}
where $\Lambda_\mu^\nu$ is a Lorentz transformation and $ \det\frac{d\Phi_I}{d\Phi_J}$ a rescaling of the volume.  Physically, $\Lambda_\mu^\nu$ moves between the frame $d\Sigma_{rest}^\mu=d\Phi_1d\Phi_2d\Phi_3(1,\vec{0})$.

We can now try define an effective Lagrangian that ``automatically'' solves Eq. \ref{ergdeltashift} at every time step, thereby guaranteeing Eq. \ref{ergdelta} holds in every frame.   Good coarse-graining variables are generators of the transformation
\begin{equation}
L(\phi) \simeq L_{eff}\left( a_\mu(\Sigma),\psi(\Sigma) \right) \eqcomma   L_{eff}\left( a_\mu(\Sigma),\psi(\Sigma) \right)= L_{eff}\left( a_\mu(\Sigma'),\psi(\Sigma') \right)
  \end{equation}
With the latter equality being enforced by solving Eq. \ref{ergdeltashift}.

Concentrating on the momentum current conservation, we notice that the dependence on foliation changes of the ratio of $\Omega(...,P_\mu)$ is the same as $\delta(...,P_\mu)$, and the shift in the Lagrangian can be reduced to a local symmetry shift.

The only change possible therefore is that of the foliation inside the $\delta$-function $\delta^4( d^3\Sigma_\mu,...)$.
One can use the classic relations of $\delta-$functions
\begin{equation}
 \delta((f(x))) = \sum_i \underbrace{\frac{\delta(x_i-a_i)}{\left| f'(x_i=a_i)\right|}}_{f(a_i)=0} \eqcomma \phi_I' = \frac{\partial_\alpha \Sigma'_I}{\partial^\alpha \Sigma^J} \Phi_J 
\eqcomma  \delta^4(\Sigma_\mu) = \mathrm{det}_{\mu \nu} \left| \frac{\partial \Sigma^\mu}{\partial \Sigma^\nu}\right| \delta^4 (\Sigma'_\mu)
  \end{equation}
to see that the ratio of $ \delta^4 \left(   d\Sigma^\nu a_\alpha \partial^\alpha \left( \delta_\nu^\mu L\right)-  dP^\mu(\Sigma_0)    \right)$ in Eq. \ref{ergdeltashift} will reduce to a ratio of $\det \frac{\partial \Phi_I}{\partial \Phi_J}$.

To construct the effective lagrangian invariant under such transformations from $\Phi_I$ one therefore
\begin{itemize}
  \item Assumes $d\Omega$ is differentiable across changes between $\Sigma$ and $\Sigma'$ so $d\Omega(P^\mu,Q,\Sigma'_0)/d\Omega(P^\mu,Q,\Sigma_0)\simeq d\Sigma'_0/d\Sigma_0$,$Q$ transforms as a scalar and $P_\mu$ as a 4-vector under changes in foliation
  \item A local rescaling can add an overall normalization factor to the lagrangian, proportional to
    \begin{equation}
      \label{rescaled}
  T_0=b \frac{dL_{eff}}{db} \eqcomma b=\sqrt{\mathrm{\det}\partial_\mu \phi_I \partial^\mu \phi_J}
  \end{equation}
To prevent this one must therefore normalize $\Phi_I \rightarrow T_0^{-1} \phi_I$ and write the lagrangian in terms of $\phi_I$ (so rescalings will cancel out).
Note that this renders $\phi_I$ unitless, as in \cite{nicolis,son}.

This procedure gives us a physical intuition for the physical interpretation of $T_0$ and $b$ in terms of the coordinates $\phi_I$.   Under the rescaling, two volume cells with the same $T_0$ will have the same lagrangian density, and hence will be in equilibrium.  It is clear that $T_0$ is therefore the microscopic temperature, possibly weighted by a unitless microscopic degeneracy $g$ (as was done in  \cite{melag,hydeft}).
We will later use this to interpret $b$ and $L_{eff}$.
\end{itemize}
Once these steps are done all $d\Sigma_0/d\Sigma_0'$ and the effective Lagrangian $L_{eff}$ is invariant under transformations leaving $\det \frac{\partial \phi_I}{\partial \phi_J}$ unchanged
    \begin{equation}
  L(a_\mu,\psi) \rightarrow L_{eff}(\phi_I,\psi) \eqcomma L(\phi_I,\psi)=L_{eff}\left( \underbrace{\frac{\partial{\phi'_I}}{\partial \phi_J}}_{\det \frac{\partial{\phi'_I}}{\partial \phi_J}=1 } \phi_J,\psi  \right)
  \end{equation}
ratios of $ \delta^4 \left(   d\Sigma^\nu a_\alpha \partial^\alpha \left( \delta_\nu^\mu L\right)-  dP^\mu(\Sigma_0)    \right)$ cancel out and \ref{ergdeltashift} reduces automatically to
\begin{equation}
  \label{ergdeltashiftq}
1 =  \frac{ \delta\left( d\Sigma^\mu  \psi \partial_\mu L   - dQ(\Sigma_0)    \right)}{ \delta\left( d\Sigma_\mu'  \psi' \partial^\mu L   - dQ'(\Sigma_0')    \right)}
\end{equation}
so all coordinate diffeomorphism dependence has been eliminated.  To complemte our task, one must also  construct the symmetries allowing Eq. \ref{ergdeltashiftq} to be satisfied automatically.  We remember that 
since $d\Sigma^\mu  \partial_\mu$ is a scalar, we can evaluate it in the frame where
$  d\Sigma^\mu = d\Phi_1 d\Phi_2 d\Phi_3 \left( 1,\vec{0}\right)$, where it becomes clear that $\partial_\mu L$ will reduce to gradients of $\partial_\mu \Phi_{1,2,3}$ and $\partial_\mu \Psi$.   Since one can always choose coordinates where  $\partial_\mu \Psi=0$, imposing Eq. \ref{ergdeltashiftq} is of course equivalent to imposing invariance of $L(\psi)$ under arbitrary shifts of functions of $\Phi_I$.

Summarizing this main result of our work, for the local ergodicity condition Eq. \ref{ergdelta2} to apply for a generic foliation $\Sigma$, the effective Lagrangian $L_{eff}$, written in terms of generators of local space translations $\phi_I$ and internal symmetry phases $\psi$, must be invariant under any $\phi_I \rightarrow \phi'_I$ leaving $\det_{IJ} \frac{\partial \phi'_I}{\partial \phi_J}$ and $\psi \rightarrow \psi+f(\phi_I)$ invariant.
These of coarse are exactly the symmetries in \cite{jackiw,nicolis,son}.

The identification of conserved currents in terms of $L_{eff}$ was done in \cite{son} by the guaranteed existence, under the above set of diffeomorphisms of the killing 4-vector defined by
\[\  u_\mu u^\mu=-1 \eqcomma u_\mu d\Sigma^\mu=u_\mu \partial^\mu \Phi_{I=1,2,3}=0   \]
which defines the boost from a generic $d\Sigma_\mu$ to one where it is of the form $d\phi_1d\phi_2d\phi_3(1,\vec{0})$ as well as two scalars, under diffeomorphism invariance, $b$ introduced in \eqref{rescaled} and $y=u_\mu \partial^\mu \psi$.   The latter is related to the Noether current \eqref{qdef} via \cite{son} the conserved charge current $J_Q^\mu=u_\mu dL/dy$.  Thus, the conservation equations defining ideal hydrodynamics, the simplest equations of an infinite family of currents determined by arbitrary $f(\phi_I)$
\begin{equation}
  \label{defentropy}
\partial_\mu s^\mu = \partial_\mu J^\mu_Q = \partial_\mu \left( f(\phi_I) J_Q^\mu \right)=0 \eqcomma s^\mu = b u^\mu 
\end{equation}
follow from local ergodicity via the symmetries already developed in \cite{son}
(Note that Kelvin's theorem is a consequence \cite{nicolis}).
This confirms the identification of $b$ with the microscopic entropy and, via basic thermodynamic relations, $L_{eff}$ with the microscopic energy density.

Physically the point here is that in the ideal hydrodynamic limit with no fluctuations one can perform a spacetime foliation \cite{zubecc} proportional to the 4-vector of lagrange multipliers under which entropy is maximized, $u_\mu/T$, which must define the direction of every conserved current.
Rotations and volume-preserving rescalings represent the residual symmetries after this vector is fixed.

While this ideal hydrodynamics is the same as the usual one \cite{kodama,rom,denicol}, this way of arriving at the ideal hydrodynamics {\em as a limit} is profoundly different from the usual approach of coarse-graining Boltzmann-type equations around conservation laws, one perhaps more useful to study hydrodynamics in small systems such as droplets of quark-gluon plasma, \cite{zajc} ultracold atoms \cite{ultracold} and perhaps the everyday phenomenon of ``the Brazilian nut effect'' \cite{braznut}.

In the usual approach \cite{denicol} hydrodynamics is a limit of a deterministic transport equation valid in the ensemble average limit.   This means a hierarchy
\begin{equation}
  L_{micro} \ll \underbrace{L_{diss}}_{\sim \eta/(sT)} \ll \underbrace{L_{macro}}_{\sim (\partial u_\mu)^{-1}}
\end{equation}
where $L_{micro}$, the origin of stochasticity is neglected and the effective theory is obtained by a coarse-graining around $L_{diss}/L_{macro}$ ($L_{diss}$ determines mean free path, or viscosity to entropy ratio $\eta/s$).

While the $L_{micro}$ expansion was not studied as extensively as the gradient expansion, it is possible to include it via diagrammatic techniques \cite{sticky} as well as recently developed approaches based on the Schwinger-Keldysh formalism \cite{sk1,steph}.    In these approaches, one can obtain an expansion in $L_{micro}/L_{diss}$ \cite{ll,gale,csernai,kovfluct,kapusta,teaney,jorgefluct,sticky,sk1,sk2,steph} as well as a renormalization of dissipative coefficients \cite{sticky} due
to contributions of fluctuations.    Unsurprisingly, this approach almost invariably results in $L_{diss}$ being
renormalized to $\geq \order{L_{micro}}$ even if the "bare" $L_{diss}$ is small.
This is a reasonable conclusion, but small systems hydrodynamics \cite{zajc,ultracold,braznut} seems to contradict this conclusion as the goodness of hydrodynamic expansions is remarkably independent on $L_{micro}$, suggesting that perhaps hydrodynamics is somehow non-analytic in $L_{micro}/L_{diss}$, as argued in \cite{bdnk}.

In the approach taken in this work, we start with the ergodic hypothesis to hold approximately in every cell at a scale $\Delta t$
, but we do not consider fluctuations to be necessarily small; In fact, the deterministic continuum is coincident with the onset of the regression to the mean by the law of large numbers.
Such ``coarse graining around ergodicity'' in the ideal limit means
\begin{equation}
  \label{ourhyerarchy}
L_{diss}=0 \eqcomma L_{micro} \equiv \Delta \ll L_{macro}
\end{equation}
If one could move beyond $L_{diss}=0$ this 
could lead To a limit to hydrodynamics that depends very differently on the thermal and mean free path scale.
The main point here is that
\begin{description}
\item[(i)] a strongly coupled system could be "close the ergodic limit" at every cell even if the number of dofs in each cell is small.   In this case the ``Khinchin conditions'' \cite{khinchin} are not satisfied, so Gibbsian and Boltzmannian equilibria are very different, but the system will be in a local Gibbsian equilibrium at every point
\item[(ii)] Since ergodicity and frame Independence are different concepts, in such a limit, foliation independence should still be valid.  It is in fact far from clear that, provided the system is strongly coupled and chaotic, decreasing the number of DoFs decreases the applicability of the ergodic hypothesis.
  In this sense, fluctuations should respect foliation independence.   Approximate foliation independence means fluctuations are inherently non-linear and in fact often help the system thermalize.  This is particularly appealing if fluctuation-dissipation holds \cite{bonanca}.   
 \end{description}
To expand on point \textbf{(i)}, the considerations above are particularly relevant when the system is small enough for quantum effects to be relevant.
In fact, One can use the direct connection between the Eigenstate thermalization hypothesis \cite{eth} (or similarly  ``Berry's conjecture, see \cite{beceth} and references therein) and the microcanonical density matrix to generalize the derivation here to a quantum system.   
A generic mixed density matrix commuting with the Hamiltonian of a theory with a conserved charge and energy-momentum is characterized by the density matrix
\begin{equation}
\label{ethdens}
\hat{\rho}=\delta_{P,P'}\delta_{Q,Q'} c_{P,Q} c_{P',Q'}^*|i,P_\mu,Q><j,P_\mu,Q| \eqcomma \hat{U}^{-1} \hat{\rho} \hat{U} = \hat{\rho}
\end{equation}
where
\begin{equation}
\hat{U}=  \exp \left[ i \hat{P}^{\nu} \delta {x}_{\nu} \right] \exp\left[i \theta \delta \hat{Q} \right]|
\end{equation}
  the phase invariance $\theta$ in a first quantized theory just reflects unitarity and $J_{\mu \nu}$ are simply generators of momentum and angular momentum.
For a highly mixed thermalized system, $\rho$ becomes a pseudo-random matrix, averaged to $ c_{P,Q} c_{P',Q'}^* \simeq 1/d\Sigma(P,Q)$, making it indistinguishable from the microcanonical ensemble examined in this paper.

Applying the Eigenstate thermalization hypothesis to every cell in every foliation is equivalent to promoting $J_{\mu \nu},\theta,P,Q$ to functions of $x_\mu$ and imposing foliation independence on the ``pseudo-randomness'' of $\hat{\rho}$.
\begin{equation}
  \left. \frac{d\hat{\rho}}{d\Sigma_0}\right|_{\Sigma_0-\Sigma_0'\simeq \Delta t}=0 \eqcomma  \hat{\rho} \simeq \frac{1}{d\Sigma} \hat{\delta}_{E,E'}\hat{\delta}_{Q,Q'}
\end{equation}
and also
\begin{equation}
  \hat{U}^{-1}(x) \hat{\rho} \hat{U}(x) \simeq \hat{\rho} \eqcomma \hat{U}(x)= \exp \left[ i \hat{T}^{\mu \nu} d^3 \Sigma_\mu \delta {x}_{\nu} \right] \exp\left[i \partial_{\alpha} \theta d^3 \Sigma^\alpha \delta \hat{Q} \right]
\end{equation}
for arbitrary $d^3\Sigma_\mu$.   By imposing $\Sigma$-invariance on the first equation it is not too difficult to see that, from \eqref{ergfol} with the density matrix as an ``observable'' an equation equivalent to Eq. \ref{ergdeltashift} can be derived, and hence quantum dynamics will be equivalent to the classical microcanonical one.

What this reasoning suggests is that if the system can be divided into cells where the Eigenstate thermalization hypothesis applies, then a dynamics equivalent to ideal fluid dynamics will naturally emerge even if the number of degrees of freedom is comparatively low and fluctuations are big.   In the context of heavy ion physics, therefore, the applicability of the Eigenstate thermalization hypothesis suggested in \cite{beceth} should go hand in hand in the onset of  hydrodynamic behavior where spacial anisotropy is also present.   Whether the Eigenstate thermalization hypothesis can be applied to QCD in the strongly coupled regime of course remains to be understood \cite{quasi}, but perhaps the parton desentanglement mechanism suggested in \cite{kharzeev} can generate local ergodicity. To verify this the diffeomorphisms examined here must be examined on the lightcone.

Regarding point \textbf{(ii)}, %on the previous page,
to try to relax the $L_{diss}=0$ limit of Eq. \ref{ourhyerarchy},
let us now imagine the system is {\em approximately} locally ergodic, defined by the idea that deviations from ergodicity are within an standard fluctuation.    $\phi_I$ can then be thought of not as coordinates but as probability distributions, each characterized by a killing vector $u_\mu$.   One heuristically arrive at the picture in \cite{bdnk}.  To do so one can use escort distributions on top of the ergodic distribution \cite{escort}
\begin{equation}
\phi_I \rightarrow \hat{\phi}_I \equiv \exp\left[ \phi_I- \frac{d \lnz}{d\Sigma_I} \right] \eqcomma Z= \mathrm{Tr} \exp \left[ d\Sigma_\mu \beta_\nu \hat{T}^{\mu \nu} \right]
\end{equation}
One can then link lagrangian hydrodynamics to Zubarev relativistic hydrodynamics as developed in \cite{zubecc}.

In addition let us suppose in accordance to \textbf{(ii)} that the fluctuation-dissipation theorem still applies, so fluctuations around equilibrium give a good estimate of deviations from ergodicity.
Supposing fluctuation-dissipation applies is equivalent to supposing that while $b$ as defined in Eq. \ref{defentropy} is not conserved, the probability of a cell evolving from a starting configuration at $b=b_0, \tau=\tau_0$ to $b=b_0+db$ at $\tau=d\tau=d\Sigma_\mu d\Sigma^\mu$ is
\begin{equation}
 \frac{P\left(\phi_I=\phi_{I0}+d\phi_I,\tau=\tau_0+d\tau | \phi_I=\phi_{I0},\tau=\tau_0\right)}{P\left(\phi_I=\phi_{I0},\tau=\tau_0+d\tau | \phi_I=\phi_{I0}+d\phi_I,\tau=\tau_0\right)} = \frac{f(b+db)}{f(b)}
\end{equation}
i.e. a function of the entropy only.   
one can easily check that, while this dynamics need not be compatible with Eq. \ref{ergdelta}, it remains invariant under foliation changes.
 Keeping the idea that the number of miscrostates is $\sim \exp(b)$ forces $f(b)$ to be an exponential as well.
 It is in fact the dynamics of \cite{zubarev} expressed in Lagrangian hydrodynamics.    The multiple definitions of $\phi_I$ should therefore accommodate the gauge-like symmetry of \cite{zubarev} described in \cite{bdnk}. Here, it is worth pointing out that if one uses local ergodicity to describe microscopic fluctuations, it becomes obvious that, as pointed out in \cite{bdnk}, only fluctuations of the total $T_{\mu \nu}$ are physical, while fluctuations of $u_\mu,e,p,\Pi_{\mu \nu}$ on their own are essentially equivalent to constrained gauge redundancies, ensembles cannot depend on them individually.   In the lagrangian picture this is evident from the fact that \cite{nicolis,son} all perturbative degrees of freedom are Goldstone modes of the symmetries emergent from ergodicity.  Distinguishing a non-hydrodynamic excitation of $\Pi_{\mu \nu}$ from a fluctuation in equilibrium $u_\mu,e$ is only possible when the the ergodic limit is exact in every cell as well as in the ensemble average, i.e. when the number of microscopic DoFs is infinite in every coarse-grained cell.

 It is also worth pointing out that any long-range potential between degrees of freedom will, by the fluctuation dissipation theorem applied to the microcanonical ensemble \cite{bonanca} be indistinguishable from an anisotropic collective flow in it's sampling dynamics.   This leads exactly to the picture advocated around eq. (36) \cite{bdnk}, where a large class of deformations of the phase space function has the potential to make Vlasov-type and Boltzmann-type terms in kinetic theory cancel out, leading to an ensemble of equivalent local equilibrium states.
 Superficially this looks like the plasma instability scenario \cite{mrow}, but the ``equivalent equilibrium'' states are really indistinguishable from bona fide equilibrium.  One can see this by considering that if local ergodicity  is approximately respected in each cell as well as across ensemble averages, hydrodynamic signatures should be expected to translate across fluctuations,is their fluctuation should reflect initial state fluctuations, as indeed seems to be the case \cite{kodama,vogel}.  If hydrodynamical observables show thermalization driven by plasma instabilities,  one would expect parametrically larger fluctuations \cite{shur}. 
 % A systematic account as to how this works out in an approximately exact ergodic hypothesis plus corrections will be left to further work.

The above reasoning assumes that the potentials are local and symmetric enough to not generate long range correlations. The alternative of course are "solids" and "jellies" \cite{nicolis,solid1,solid2,solid3}, where long range correlations break local deformation, isotropy and translation symmetries.   
Looking at the arguments governing the applicability of the ergodic hypothesis \cite{khinchin}, it is clear that 
while such systems have an equilibrium state characterized by the local maximum of entropy, their accessible phase space volume is generally far away from the indecomposable limit \cite{khinchin}, because of long range correlations imposing constraints local to phase space.   Thus, local ergodicity in the sense of this work does not apply to these systems.  Moreover, the arguments presented here make apparent why such systems are generally {\em fragile}:  The lack of phase space indecomposability \cite{khinchin} ensures that small perturbations will lead to large deviations from the ergodic hypothesis in every cell, and hence it is expceted that it becomes relatively easy to "break" such systems, i.e. bring them far away from local equilibrium.   The preparation of such states will require a careful cancellation of dynamics at different scales. Thus, relativistic fluids and the symmetries defining them are unique in that local ergodicity ensures both the existence of continuous deformation symmetries and the fact that adjacent microstates to equilibrium will be sampled, by local perturbations, in a way that guarantees the continuation of near-local equilibrium.

 Making a link of this work with hydrodynamics with spin \cite{lisa,hydeft} is a non-trivial endeavor, because of the fact that the interplay of spin with angular momentum lead to non-local phase space hypersurfaces with a high deviation from indecomposability \cite{khinchin} (the presence of subregions of phase space with limited connection can be interpreted as the appearance of an intermediate scale \cite{kaminski}).   Nevertheless, it is possible that hydrodynamics with spin can {\em only} be defined this way, given the fact that spin-vorticity coupling alters microstate distributions at the same order as hydrodynamic fluctuations \cite{hydeft}, something arising as non-locality from transport models \cite{nora}.

 Finally, as a wild speculation taking inspiration from  \cite{jacobson}, the addition of a horizon term to the entropy, $d\Omega$, and a generalization of the ergodicity hypothesis  from time averages to worldline/proper time averages might open the way to an enthropic gravity scenario where the equivalence principle is respected exactly, as suggested in \cite{zubarev,deco}.

 In conclusion we have shown that the local symmetries associated with ideal lagrangian hydrodynamics can be thought of as emerging from a hypothesis of local ergodicity strong enough to be invariant under smooth spacetime foliations.   This opens a way of thinking about hydrodynamics which is explicitly ''non-perturbative'' with respect to fluctuations and more appropriate to study the onset of hydrodynamic behavior in small systems.

 \textbf{Acknowledgements}
 GT thanks CNPQ bolsa de produtividade 305731/2023-8 and bolsa FAPESP  2023/06278-2.
 I also thank Dam Thanh Son for providing a key insight that led to this development during his visit to Unicamp.  I also thank Marcus Segantini Bonanca for discussions and suggestions
  

\begin{thebibliography}{10}

\bibitem{kodama}
R.~Derradi de Souza, T.~Koide and T.~Kodama,
%``Hydrodynamic Approaches in Relativistic Heavy Ion Reactions,''
Prog. Part. Nucl. Phys. \textbf{86} (2016), 35-85
doi:10.1016/j.ppnp.2015.09.002
[arXiv:1506.03863 [nucl-th]].

\bibitem{rom}
P.~Romatschke and U.~Romatschke,
%``Relativistic Fluid Dynamics In and Out of Equilibrium,''
Cambridge University Press, 2019,
ISBN 978-1-108-48368-1, 978-1-108-75002-8
doi:10.1017/9781108651998
[arXiv:1712.05815 [nucl-th]].

\bibitem{heinz}
S.~Jeon and U.~Heinz,
%``Introduction to Hydrodynamics,''
Int. J. Mod. Phys. E \textbf{24} (2015) no.10, 1530010
doi:10.1142/S0218301315300106
[arXiv:1503.03931 [hep-ph]].

  \bibitem{denicol} G.Denicol,D.Rischke ``Microscopic Foundations of Relativistic Fluid Dynamics'', Springer(2021)

    \bibitem{hydeft}
D.~Montenegro, R.~Ryblewski and G.~Torrieri,
%``Relativistic fluid dynamics and its extensions as an effective field theory,''
Acta Phys. Polon. B \textbf{50} (2019), 1275
doi:10.5506/APhysPolB.50.1275
[arXiv:1903.08729 [hep-th]].


\bibitem{zubecc}
F.~Becattini, M.~Buzzegoli and E.~Grossi,
%``Reworking the Zubarev's approach to non-equilibrium quantum statistical mech>
Particles \textbf{2} (2019) no.2, 197-207
doi:10.3390/particles2020014
[arXiv:1902.01089 [cond-mat.stat-mech]].
%17 citations counted in INSPIRE as of 10 Apr 2021



\bibitem{zajc}
J.~L.~Nagle and W.~A.~Zajc,
%``Small System Collectivity in Relativistic Hadronic and Nuclear Collisions,''
Ann. Rev. Nucl. Part. Sci. \textbf{68}, 211-235 (2018)
doi:10.1146/annurev-nucl-101916-123209
[arXiv:1801.03477 [nucl-ex]].

\bibitem{ultracold}
A.~Adams, L.~D.~Carr, T.~Sch\"afer, P.~Steinberg and J.~E.~Thomas,
%``Strongly Correlated Quantum Fluids: Ultracold Quantum Gases, Quantum Chromodynamic Plasmas, and Holographic Duality,''
New J. Phys. \textbf{14} (2012), 115009
doi:10.1088/1367-2630/14/11/115009
[arXiv:1205.5180 [hep-th]].

\bibitem{braznut} A. P. J. Breu, H.-M. Ensner, C. A. Kruelle, and I. Rehberg,Phys.Rev.Lett. {\bf 90} 014302-1 (2003)

\bibitem{attractor}
G.~Giacalone, A.~Mazeliauskas and S.~Schlichting,
%``Hydrodynamic attractors, initial state energy and particle production in relativistic nuclear collisions,''
Phys. Rev. Lett. \textbf{123} (2019) no.26, 262301
doi:10.1103/PhysRevLett.123.262301
[arXiv:1908.02866 [hep-ph]].


\bibitem{mrow}
S.~Mrowczynski, B.~Schenke and M.~Strickland,
%``Color instabilities in the quark\textendash{}gluon plasma,''
Phys. Rept. \textbf{682} (2017), 1-97
doi:10.1016/j.physrep.2017.03.003
[arXiv:1603.08946 [hep-ph]].


\bibitem{khinchin}
  A.Y.Khinchin,''Mathematical foundations of Statistical Mechanics''

  
\bibitem{jaynes} E. T. Jaynes,
  American Journal of physics \textbf{33} 391-398 (1965)
  doi:10.1119/1.1971557

\bibitem{ll} Lifshitz and Landau, Statistical Mechanics II

\bibitem{csernai}
L.~P.~Csernai, S.~Jeon and J.~I.~Kapusta,
%``Fluctuation and dissipation in classical many particle systems,''
Phys. Rev. A \textbf{56} (1997), 6668
doi:10.1103/PhysRevE.56.6668
[arXiv:nucl-th/9708033 [nucl-th]].

\bibitem{gale}
M.~Singh, C.~Shen, S.~McDonald, S.~Jeon and C.~Gale,
%``Hydrodynamic Fluctuations in Relativistic Heavy-Ion Collisions,''
Nucl. Phys. A \textbf{982} (2019), 319-322
doi:10.1016/j.nuclphysa.2018.10.061
[arXiv:1807.05451 [nucl-th]].

\bibitem{kapusta}
J.~I.~Kapusta, B.~Muller and M.~Stephanov,
%``Relativistic Theory of Hydrodynamic Fluctuations with Applications to Heavy Ion Collisions,''
Phys. Rev. C \textbf{85} (2012), 054906
doi:10.1103/PhysRevC.85.054906
[arXiv:1112.6405 [nucl-th]].

\bibitem{teaney}
Y.~Akamatsu, A.~Mazeliauskas and D.~Teaney,
%``A kinetic regime of hydrodynamic fluctuations and long time tails for a Bjorken expansion,''
Phys. Rev. C \textbf{95} (2017) no.1, 014909
doi:10.1103/PhysRevC.95.014909
[arXiv:1606.07742 [nucl-th]].

%\cite{Kovtun:2012rj}
\bibitem{kovfluct}
P.~Kovtun,
%``Lectures on hydrodynamic fluctuations in relativistic theories,''
J. Phys. A \textbf{45} (2012), 473001
doi:10.1088/1751-8113/45/47/473001
[arXiv:1205.5040 [hep-th]].


\bibitem{jorgefluct}
N.~Mullins, M.~Hippert and J.~Noronha,
%``Stochastic fluctuations in relativistic fluids: causality, stability, and the information current,''
[arXiv:2306.08635 [nucl-th]].


\bibitem{sticky}
P.~Kovtun, G.~D.~Moore and P.~Romatschke,
%``The stickiness of sound: An absolute lower limit on viscosity and the breakdown of second order relativistic hydrodynamics,''
Phys. Rev. D \textbf{84} (2011), 025006
doi:10.1103/PhysRevD.84.025006
[arXiv:1104.1586 [hep-ph]].

\bibitem{sk1}
H.~Liu and P.~Glorioso,
%``Lectures on non-equilibrium effective field theories and fluctuating hydrodynamics,''
PoS \textbf{TASI2017} (2018), 008
doi:10.22323/1.305.0008
[arXiv:1805.09331 [hep-th]].

\bibitem{sk2}
A.~Jain and P.~Kovtun,
%``Schwinger-Keldysh effective field theory for stable and causal relativistic hydrodynamics,''
[arXiv:2309.00511 [hep-th]].

\bibitem{steph}
X.~An, G.~Basar, M.~Stephanov and H.~U.~Yee,
%``Relativistic Hydrodynamic Fluctuations,''
Phys. Rev. C \textbf{100} (2019) no.2, 024910
doi:10.1103/PhysRevC.100.024910
[arXiv:1902.09517 [hep-th]].

%\cite{Torrieri:2020ezm}
\bibitem{zubarev}
G.~Torrieri,
%``Fluctuating Relativistic hydrodynamics from Crooks theorem,''
JHEP \textbf{02} (2021), 175
doi:10.1007/JHEP02(2021)175
[arXiv:2007.09224 [hep-th]].
%7 citations counted in INSPIRE as of 21 Jun 2023

\bibitem{bonanca}
M.Bonanca, 
Phys. Rev. E \textbf{78}, 031107 (2008)
[arXiv: 0804.0107 [cond-mat]].

\bibitem{kovtun}
R.~E.~Hoult and P.~Kovtun,
%``Stable and causal relativistic Navier-Stokes equations,''
JHEP \textbf{06} (2020), 067
doi:10.1007/JHEP06(2020)067
[arXiv:2004.04102 [hep-th]].

\bibitem{prx}
F.~S.~Bemfica, M.~M.~Disconzi and J.~Noronha,
%``First-Order General-Relativistic Viscous Fluid Dynamics,''
Phys. Rev. X \textbf{12} (2022) no.2, 021044
doi:10.1103/PhysRevX.12.021044
[arXiv:2009.11388 [gr-qc]].

\bibitem{spalinski}
J.~Noronha, M.~Spali\'nski and E.~Speranza,
%``Transient Relativistic Fluid Dynamics in a General Hydrodynamic Frame,''
Phys. Rev. Lett. \textbf{128} (2022) no.25, 252302
doi:10.1103/PhysRevLett.128.252302
[arXiv:2105.01034 [nucl-th]].


\bibitem{gavassino}
L.~Gavassino, M.~Antonelli and B.~Haskell,
%``Thermodynamic Stability Implies Causality,''
Phys. Rev. Lett. \textbf{128} (2022) no.1, 010606
doi:10.1103/PhysRevLett.128.010606
[arXiv:2105.14621 [gr-qc]].

\bibitem{gava}
L.~Gavassino, 
%``The zeroth law of thermodynamics in special relativity,''
Found. Phys. \textbf{50} (2020) no.11, 1554-1586
doi:10.1007/s10701-020-00393-x
[arXiv:2005.06396 [gr-qc]].


\bibitem{bdnk}
T.~Dore, L.~Gavassino, D.~Montenegro, M.~Shokri and G.~Torrieri,
%``Fluctuating relativistic dissipative hydrodynamics as a gauge theory,''
[arXiv:2109.06389 [hep-th]].


\bibitem{cf}
F.~Cooper and G.~Frye,
%``Comment on the Single Particle Distribution in the Hydrodynamic and Statistical Thermodynamic Models of Multiparticle Production,''
Phys. Rev. D \textbf{10} (1974), 186
doi:10.1103/PhysRevD.10.186

\bibitem{grfol}
E.~Gourgoulhon,
%``3+1 formalism and bases of numerical relativity,''
[arXiv:gr-qc/0703035 [gr-qc]].


\bibitem{arnold}
P.~B.~Arnold,
%``Quark-Gluon Plasmas and Thermalization,''
Int. J. Mod. Phys. E \textbf{16} (2007), 2555-2594
doi:10.1142/S021830130700832X
[arXiv:0708.0812 [hep-ph]].
%61 citations counted in INSPIRE as of 11 Jul 2022

  %\cite{Gelis:2019yfm}
\bibitem{gelis}
F.~Gelis,``Quantum Field Theory,'',ISBN-13: 978-1108480901\\
\url{https://www.ipht.fr/Pisp/francois.gelis/Physics/2018-QFT.pdf}

\bibitem{calzetta}
E.~A.~Calzetta and B.~L.~B.~Hu,
%``Nonequilibrium Quantum Field Theory,''
Oxford University Press, 2009,
ISBN 978-1-00-929003-6, 978-1-00-928998-6, 978-1-00-929002-9, 978-0-511-42147-1, 978-0-521-64168-5
doi:10.1017/9781009290036


\bibitem{jackiw}
R.~Jackiw, V.~P.~Nair, S.~Y.~Pi and A.~P.~Polychronakos,
%``Perfect fluid theory and its extensions,''
J. Phys. A \textbf{37} (2004), R327-R432
doi:10.1088/0305-4470/37/42/R01
[arXiv:hep-ph/0407101 [hep-ph]].
  
\bibitem{nicolis}
S.~Dubovsky, T.~Gregoire, A.~Nicolis and R.~Rattazzi,
%``Null energy condition and superluminal propagation,''
JHEP \textbf{03} (2006), 025
doi:10.1088/1126-6708/2006/03/025
[arXiv:hep-th/0512260 [hep-th]].
  
  %\bibitem{Dubovsky:2011sj} 
\bibitem{son}
  S.~Dubovsky, L.~Hui, A.~Nicolis and D.~T.~Son,
  %``Effective field theory for hydrodynamics: thermodynamics, and the
  %derivative expansion,'' 
  Phys.\ Rev.\ D {\bf 85}, 085029 (2012)
%  [arXiv:1107.0731 [hep-th]].
  %%CITATION = ARXIV:1107.0731;%%


  \bibitem{analogue}
C.~Barcelo, S.~Liberati and M.~Visser,
%``Analogue gravity,''
Living Rev. Rel. \textbf{8} (2005), 12
doi:10.12942/lrr-2005-12
[arXiv:gr-qc/0505065 [gr-qc]].


\bibitem{jacobson}
T.~Jacobson,
%``Thermodynamics of space-time: The Einstein equation of state,''
Phys. Rev. Lett. \textbf{75} (1995), 1260-1263
doi:10.1103/PhysRevLett.75.1260
[arXiv:gr-qc/9504004 [gr-qc]].

\bibitem{deco}
G.~Torrieri,
%``The equivalence principle and inertial-gravitational decoherence,''
[arXiv:2210.08586 [gr-qc]].

\bibitem{melag}
G.~Torrieri,
%``Viscosity of An Ideal Relativistic Quantum Fluid: A Perturbative study,''
Phys. Rev. D \textbf{85} (2012), 065006
doi:10.1103/PhysRevD.85.065006
[arXiv:1112.4086 [hep-th]].

\bibitem{eth}
L.~D'Alessio, Y.~Kafri, A.~Polkovnikov and M.~Rigol,
%``From quantum chaos and eigenstate thermalization to statistical mechanics and thermodynamics,''
Adv. Phys. \textbf{65} (2016) no.3, 239-362
doi:10.1080/00018732.2016.1198134
[arXiv:1509.06411 [cond-mat.stat-mech]].

\bibitem{beceth}
F.~Becattini,
%``An Introduction to the Statistical Hadronization Model,''
[arXiv:0901.3643 [hep-ph]].


\bibitem{quasi}
M.~Srdinsek, T.~Prosen and S.~Sotiriadis,
%``Ergodicity breaking and deviation from Eigenstate Thermalisation in relativistic QFT,''
[arXiv:2303.15123 [cond-mat.stat-mech]].


\bibitem{kharzeev}
D.~E.~Kharzeev,
%``Quantum information approach to high energy interactions,''
Phil. Trans. A. Math. Phys. Eng. Sci. \textbf{380} (2021) no.2216, 20210063
doi:10.1098/rsta.2021.0063
[arXiv:2108.08792 [hep-ph]].


\bibitem{escort}Beck, C.; Schlögl, F. Thermodynamics of Chaotic Systems; Cambridge University Press: New York, NY, USA, 1993




\bibitem{vogel}
S.~Vogel, G.~Torrieri and M.~Bleicher,
%``Elliptic flow fluctuations in heavy ion collisions at RHIC and the perfect fluid hypothesis,''
Phys. Rev. C \textbf{82} (2010), 024908
doi:10.1103/PhysRevC.82.024908
[arXiv:nucl-th/0703031 [nucl-th]].

\bibitem{shur}
S.~Mrowczynski and E.~V.~Shuryak,
%``Elliptic flow fluctuations,''
Acta Phys. Polon. B \textbf{34} (2003), 4241-4256
[arXiv:nucl-th/0208052 [nucl-th]].

\bibitem{solid1}
M.~Baggioli and S.~Grieninger,
%``Zoology of solid \textbackslash{}\& fluid holography \textemdash{} Goldstone modes and phase relaxation,''
JHEP \textbf{10}, 235 (2019)
doi:10.1007/JHEP10(2019)235
[arXiv:1905.09488 [hep-th]].

\bibitem{solid2}
L.~Gavassino, M.~M.~Disconzi and J.~Noronha,
%``Universality Classes of Relativistic Fluid Dynamics II: Applications,''
[arXiv:2302.05332 [nucl-th]].

\bibitem{solid3}
A.~Nicolis, R.~Penco, F.~Piazza and R.~Rattazzi,
%``Zoology of condensed matter: Framids, ordinary stuff, extra-ordinary stuff,''
JHEP \textbf{06}, 155 (2015)
doi:10.1007/JHEP06(2015)155
[arXiv:1501.03845 [hep-th]].

\bibitem{lisa}
F.~Becattini and M.~A.~Lisa,
%``Polarization and Vorticity in the Quark\textendash{}Gluon Plasma,''
Ann. Rev. Nucl. Part. Sci. \textbf{70} (2020), 395-423
doi:10.1146/annurev-nucl-021920-095245
[arXiv:2003.03640 [nucl-ex]].

%\cite{Hongo:2021ona}
\bibitem{kaminski}
M.~Hongo, X.~G.~Huang, M.~Kaminski, M.~Stephanov and H.~U.~Yee,
%``Relativistic spin hydrodynamics with torsion and linear response theory for spin relaxation,''
JHEP \textbf{11}, 150 (2021)
doi:10.1007/JHEP11(2021)150
[arXiv:2107.14231 [hep-th]].
%58 citations counted in INSPIRE as of 23 Jul 2023

\bibitem{nora}
N.~Weickgenannt, E.~Speranza, X.~l.~Sheng, Q.~Wang and D.~H.~Rischke,
%``Generating Spin Polarization from Vorticity through Nonlocal Collisions,''
Phys. Rev. Lett. \textbf{127} (2021) no.5, 052301
doi:10.1103/PhysRevLett.127.052301
[arXiv:2005.01506 [hep-ph]].
\end{thebibliography}
\end{document}